\documentclass[slac_one]{revtex4}
\usepackage{graphicx}
\usepackage{fancyhdr}
\usepackage{xspace}
\newcommand{\ttbar}{\ensuremath{{t\bar{t}}}\xspace}

\newcommand {\ifb}{\ensuremath{{\mathrm{fb}^{-1}}}\xspace}
\newcommand {\DO}{D\O\xspace}
\newcommand{\GeV}{\ensuremath{\mbox{GeV}}\xspace}

\newcommand{\ttH}{\ensuremath{{t\bar{t}H}}\xspace}
\newcommand{\st}{\ensuremath{\tilde{t}}\xspace}

\begin{document}

\title{Things decaying into top quarks, that are similar to top quarks, 
  or that are produced with top quarks} 

%

\author{J. Cammin on behalf of the CDF and D0 Collaborations}
\affiliation{University of Rochester, Rochester, NY 14627, USA}

\begin{abstract}
  Searches for new physics in the top-quark sector using data from
  proton-antiproton collisions at the Fermilab Tevatron are discussed.
  The large data sets collected by the D0 and CDF experiments allow
  for precision measurements of the standard model top-quark
  production rates and top quark properties so that deviations from
  the standard model expectations can be interpreted as signs of new physics. The
  presented analyses exploit the fact that the new physics would
  reveal itself in final states that are similar or identical to those
  of standard model top-antitop production.
\end{abstract}

\maketitle

\thispagestyle{fancy}


\section{INTRODUCTION} 
The production of top quark pairs at the Fermilab Tevatron provides us
with an opportunity to perform a large variety of measurements related
to the production and decay mechanisms of the top quark and many of
its properties. Given the large integrated luminosity available at the
Tevatron many of these measurements are no longer limited by
statistical uncertainties.
Consequently, it is possible to test the standard model (SM) in the
top quark sector with high accuracy and to detect deviations that
could be caused by new physics. New physics could reveal itself in
many ways, by altering the production cross section, the top quark
decay modes, the top quark mass distribution or the coupling
constants, etc. The signature of new particles can be similar to or
exactly the same as that of \ttbar pairs. This motivates to look for
signs of new physics using the same tools and analysis strategies
developed for measurements in the top quark sector. In the following
we will describe four examples: searches for particles decaying into
\ttbar, searches for the supersymmetric partners of top quarks, a
search for top-like objects $t'$, and a search for the production of a
light SM Higgs boson in association with a pair of top quarks. The
analyses are based on up to 3~\ifb of data collected by the CDF and
\DO experiments.

\section{\boldmath $t\bar{t}$ RESONANCES}
Several theories beyond the SM predict new $Z$-like particles that
decay into a pair of top-quarks, causing a peak in the \ttbar
invariant mass distribution. Such resonances could occur as
Kaluza-Klein states of the $Z$ or the gluon, in axigluon models or
theories with top color. Both CDF and \DO have searched for narrow
resonances $Z'$ assuming that the width is smaller than the detector
mass resolution~\cite{c:CDF_Zprime1, c:D0_Zprime}. Final states where
one top quark decays hadronically ($t \to Wb \to q\bar{q}'b$) and the
other one semileptonically ($t \to Wb \to \ell\nu b, \ell=e,\mu$) have
been considered (``$\ell+\text{jets}$'' final state). The CDF analysis
is based on approximately 1~\ifb of data and reconstructs the
$M_{\ttbar}$ distribution in events with at least four jets from the
jet-parton assignment that is most likely to come from \ttbar events.
The most consistent combination is found by minimizing a $\chi^2$
which is constructed from the reconstructed $W$ boson and top quark
masses, and using additional information from $b$-tagged jets. No
structure from a resonant signal is observed in the \ttbar mass
spectrum over the expected contributions from SM \ttbar and non-\ttbar
backgrounds and limits are set on the production cross section times
branching ratio $Z' \to \ttbar$ for $Z$-like resonances. For the
specific case of topcolor leptophobic models a $Z'$ with a mass below
720~GeV is ruled out at the 95\% CL.  \DO's search is based on a data
set of 2.1~\ifb in the $\ell+\text{jets}$ final state requiring at
least one $b$-tagged jet.  No attempt is made to match partons to
jets, and the invariant \ttbar mass is calculated directly from the
four-momenta of up to four leading jets, the charged lepton momentum
and the neutrino momentum.  The neutrino $p_z$ component is calculated
from the $W$ mass constraint $m_W^2 = (p_\ell + p_\nu)^2$. The
advantage of the direct reconstruction is that it allows to include
events with only three jets. This recuperates sensitivity to events
from a high mass $Z'$ where merging of two jets into a single jet due
to the relativistic boost occurs more often. With the data describing
the mass shape expected from SM background only
(Fig.~\ref{fig:D0_Zprime}), a $Z'$ mass below 760~GeV for
topcolor-assisted technicolor models is excluded at the 95\% CL.

\begin{figure}
  \begin{minipage}[t]{0.48\linewidth}
    \centering
    \includegraphics[width=\linewidth]{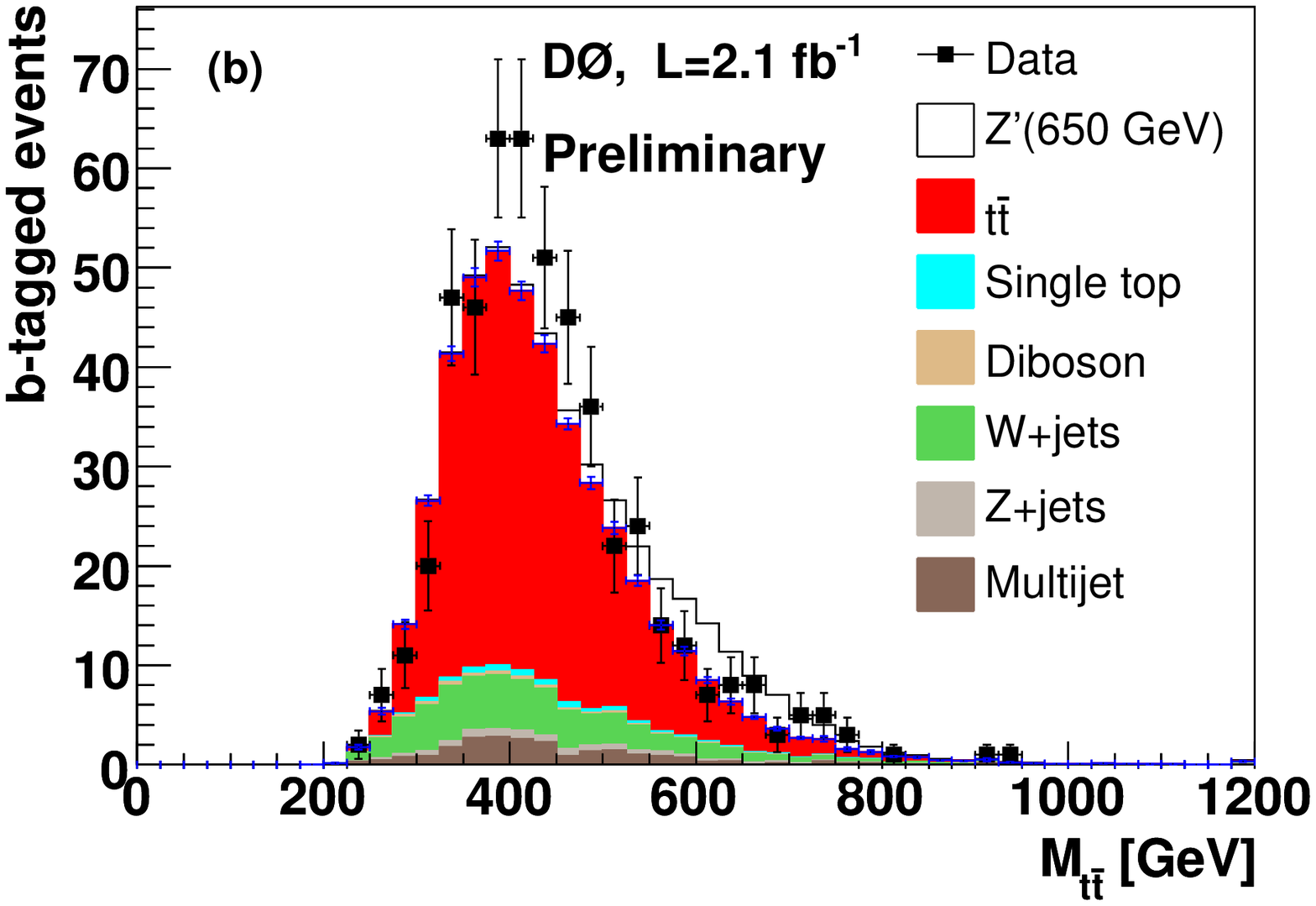}
    \caption{Distribution of the $t\bar{t}$ invariant mass in the
      $\ell+4$ or more jets channel in data and in the expected signal
      plus background. The error bars indicate the statistical error
      on the data.}
    \label{fig:D0_Zprime}
  \end{minipage}
  \hfill
  \begin{minipage}[t]{0.48\linewidth}
    \centering
    \includegraphics[width=0.7\linewidth]{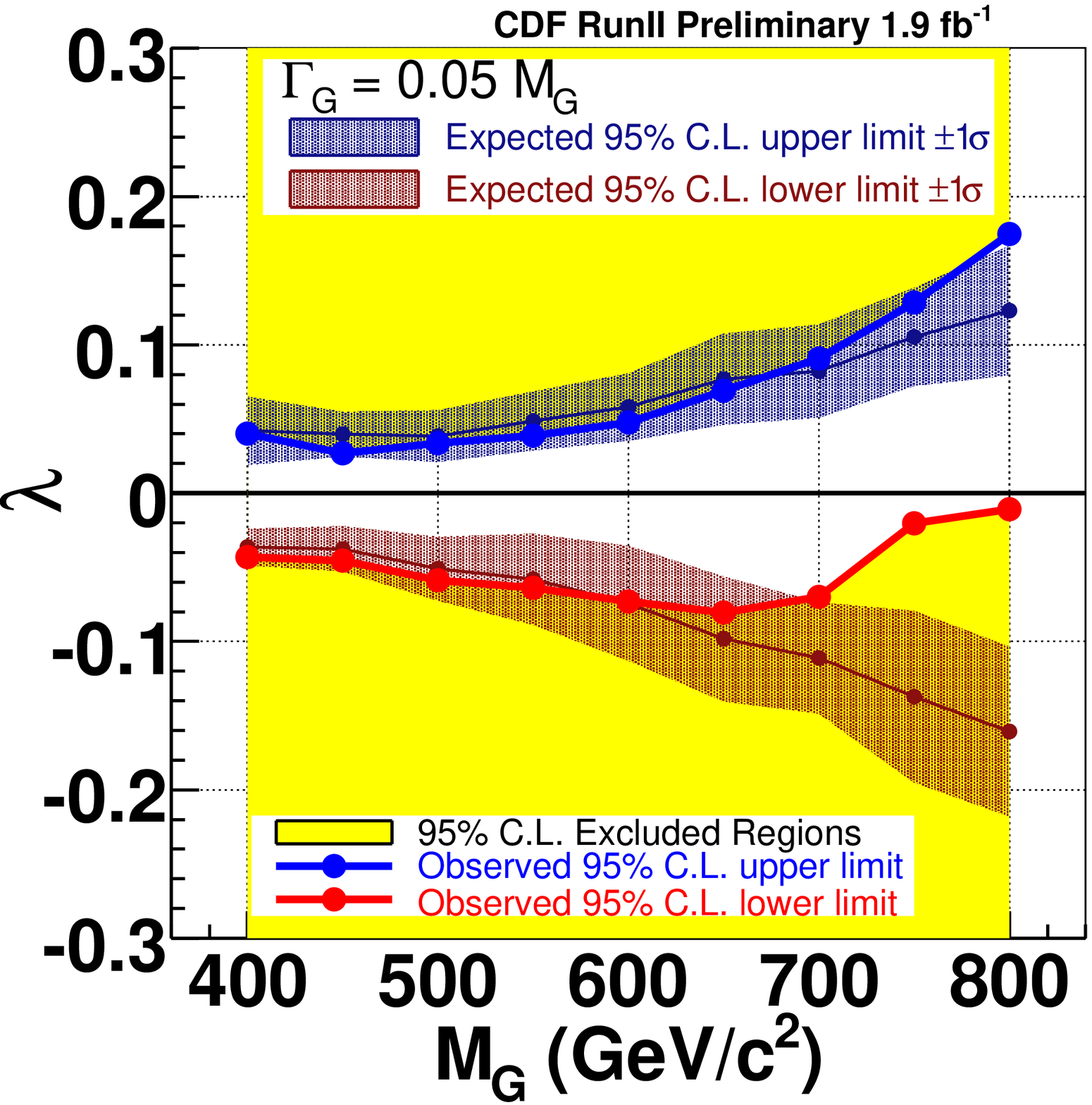}
    \caption{Expected and observed 95\% CL excluded regions in the $\lambda-M_G$ parameters space in
      the CDF search for massive gluons. This example assumes
      $\Gamma_G = 0.05\cdot M_G$.}
    \label{fig:CDF_Zprime2}
  \end{minipage}
\end{figure}

CDF has also searched for new color-octet particles (massive gluons)
that decay into \ttbar in 1.9~\ifb of data~\cite{c:CDF_Zprime2}. The
model introduces three new parameters, the coupling strength
$\lambda$, the mass $M_G$ and the decay width $\Gamma_G$ of the heavy
gluon. The $\ell+\text{jets}$ final state with at least one $b$-tagged
jet is reconstructed using the \emph{Dynamical Likelihood
  Method}~\cite{c:DLM} but omitting information from the matrix
element of the process to avoid a bias towards specific model
assumptions. Without an observed excess over the SM background, limits
are set on the coupling $\lambda$ for massive gluon widths from
$\Gamma_G = 0.05\cdot M_G$ to $\Gamma_G = 0.5\cdot M_G$
(see Fig.~\ref{fig:CDF_Zprime2} for an example).

\section{SUPERSYMMETRIC TOP QUARKS}
The decays of the supersymmetric partners of the top quark, \st or
stop quark, could lead to the same final state signature as that from
SM top quarks. Some models like electroweak baryogenesis scenarios
favor stop quarks that are lighter than the top quark. If the
superymmetric parameter space is further constrained by assuming that
$m_{\tilde{\chi}^\pm} < m_{\tilde{t}_1} - m_b$ so that the stop quark
decays exclusively into $\tilde{t}_1 \to b \chi^+_1$ and assuming that
$\tilde{\chi}_1^0$ is the LSP and that $\tilde{q}, \tilde{\ell},
\tilde{\nu}$ are heavy then the $\tilde{t}\bar{\tilde{t}}$ final
states involving $b$ quarks, one or more leptons, missing transverse
energy, and jets are identical to the SM top-pair decays in the
$\ell$+jets or dilepton final states. Hence, contributions from stop
pairs could be contained in the presumed top-quark pair data. 
CDF searched for signs of stop quarks using 2.7~\ifb of
data~\cite{c:CDF_stop} in the final state where both charginos decay
into $\tilde{\chi}^0_1 + \ell + \nu$, mimicking the \ttbar signature
in the dilepton final state. Due to the undetected two neutralinos and
two neutrinos, the stop quark mass cannot be reconstructed directly.
Instead, the mass is estimated using the neutrino weighting
method~\cite{c:CDF_NW} and the reconstructed mass is used as a
discriminant in the limit setting procedure. 
CDF sets limits on the $m_{\chi^0_1}-m_{\st}$ parameter space for
several values of the $m_{\chi^\pm_1}$ mass. An example for
$m_{\chi^\pm_1}=105.8~\GeV$ and various assumptions about the
branching ratio $\chi^\pm_1 \to \chi^0_1 \nu \ell$ is shown in
Fig.~\ref{fig:CDF_stop}.  A similar search has been performed by \DO
using 1~\ifb of data but in the $\ell$+jets final
state~\cite{c:D0_stop}. The chargino and neutralino masses are fixed
close to their experimental lower limits. Several input variables are
combined into a likelihood discriminant whose distribution is used to
extract the stop-pair signal. The 95\% CL limits on the cross section
are show in Fig.~\ref{fig:D0_stop}. The observed limits are a factor
of 7--12 higher than the theoretical prediction.

\begin{figure}
  \begin{minipage}[t]{0.48\linewidth}
    \centering
    \includegraphics[width=\linewidth]{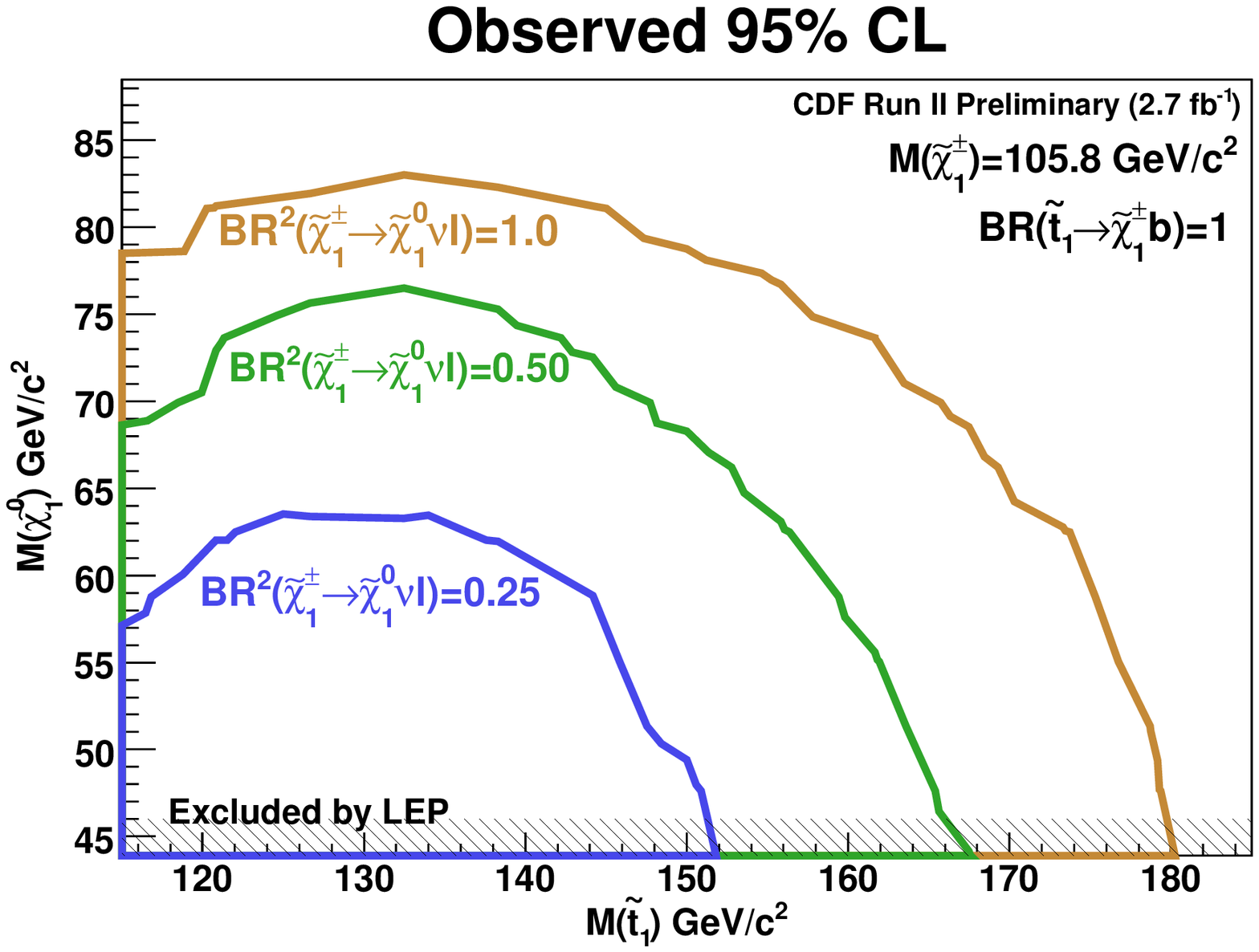}
    \caption{95\% CL exclusion regions in the
    $M(\chi^0_1)-M(\tilde{t}_1)$ parameter space for several values of
    the squared branching fraction $BR^2(\tilde{\chi}^\pm_1 \to
    \tilde{\chi}^0_1 \nu \ell)$. In this example a
    chargino mass of $M(\tilde{\chi}^\pm_1) = 105.8~\GeV$ is assumed.}
    \label{fig:CDF_stop}
  \end{minipage}
  \hfill
  \begin{minipage}[t]{0.48\linewidth}
    \centering
    \includegraphics[width=0.7\linewidth,clip,trim= 380 5 20 5 ]{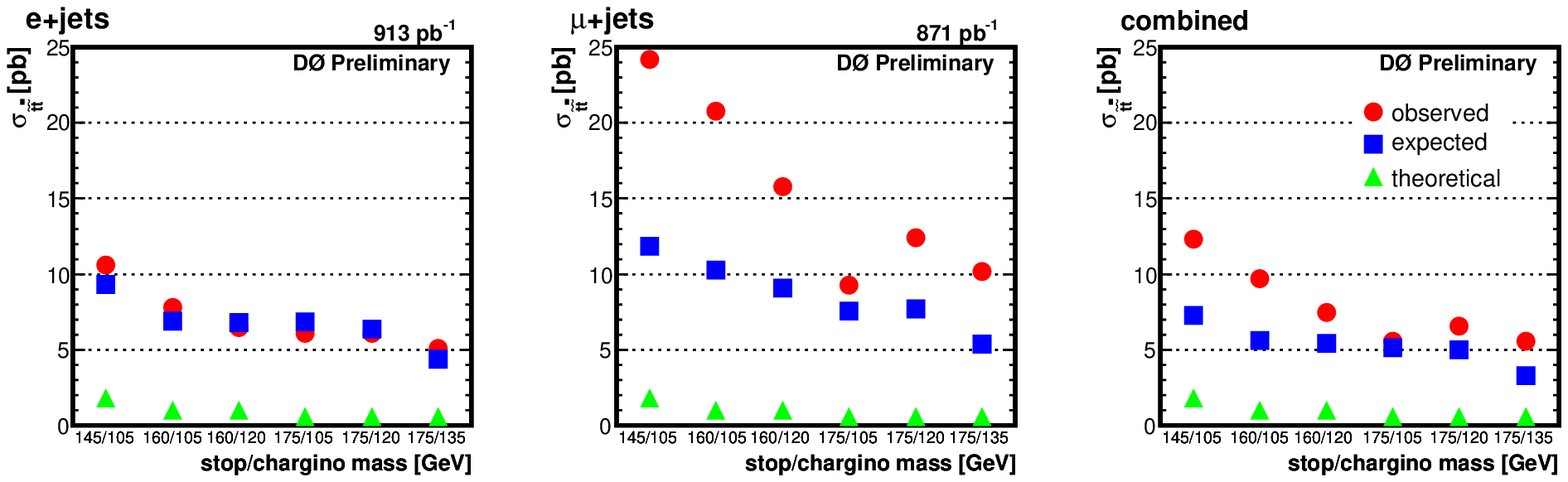}
    \caption{Expected and observed 95\% CL cross section limits on the
      production of a pair of stop quarks as a function of the stop an
      chargino mass. Also shown is the theoretically predicted cross
      section. }
    \label{fig:D0_stop}
  \end{minipage}
\end{figure}

\section{FOURTH GENERATION, TOP-LIKE OBJECTS}
The possibility of new top-like objects from a massive fourth
generation of quarks has been studied by CDF based on 2.8~\ifb of
data~\cite{c:tprime}. Such objects are predicted by various models,
for example Little Higgs or Beautiful Mirrors scenarios. Assuming that
the mass of the new $t'$ is larger than the top quark mass and that
$m_{t'}-m_{b'} < m_W$ so that $t' \to qW$ exclusively, a template fit
method based on a minimal $\chi^2$ to find the correct combination of
parton-jets assignment is used to set limits on the $t'$ mass. No
$b$-tagging is applied to avoid making assumptions about the coupling
of the $t'$ to the SM quark sector. Both the reconstructed $t'$ mass
(Fig.~\ref{fig:CDF_tprime1}) and the scalar sum $H_T$ of the
transverse momenta of all reconstructed objects in the event show good
discrimination between the signal, the \ttbar and other backgrounds. A
two-dimensional likelihood fit to these variables is applied to
extract the amount of signal that would be allowed by the data. Using
this information, limits are set on the production rate $p\bar{p} \to
t'\bar{t}'$ as a function of $m_{t'}$, see Fig~\ref{fig:CDF_tprime2}.
An analysis of the data in the high $H_T$ and reconstructed $m_{t'}$ tails of the 2D
distribution show that no significant excess of data over the
background predictions is observed.
By comparing the observed cross section limit to the expected signal
cross section, a $t'$ with a mass smaller than 311~\GeV is excluded.

\begin{figure}
  \begin{minipage}[t]{0.48\linewidth}
    \centering
    \includegraphics[width=0.8\linewidth,clip,trim= 20 20 20 35]{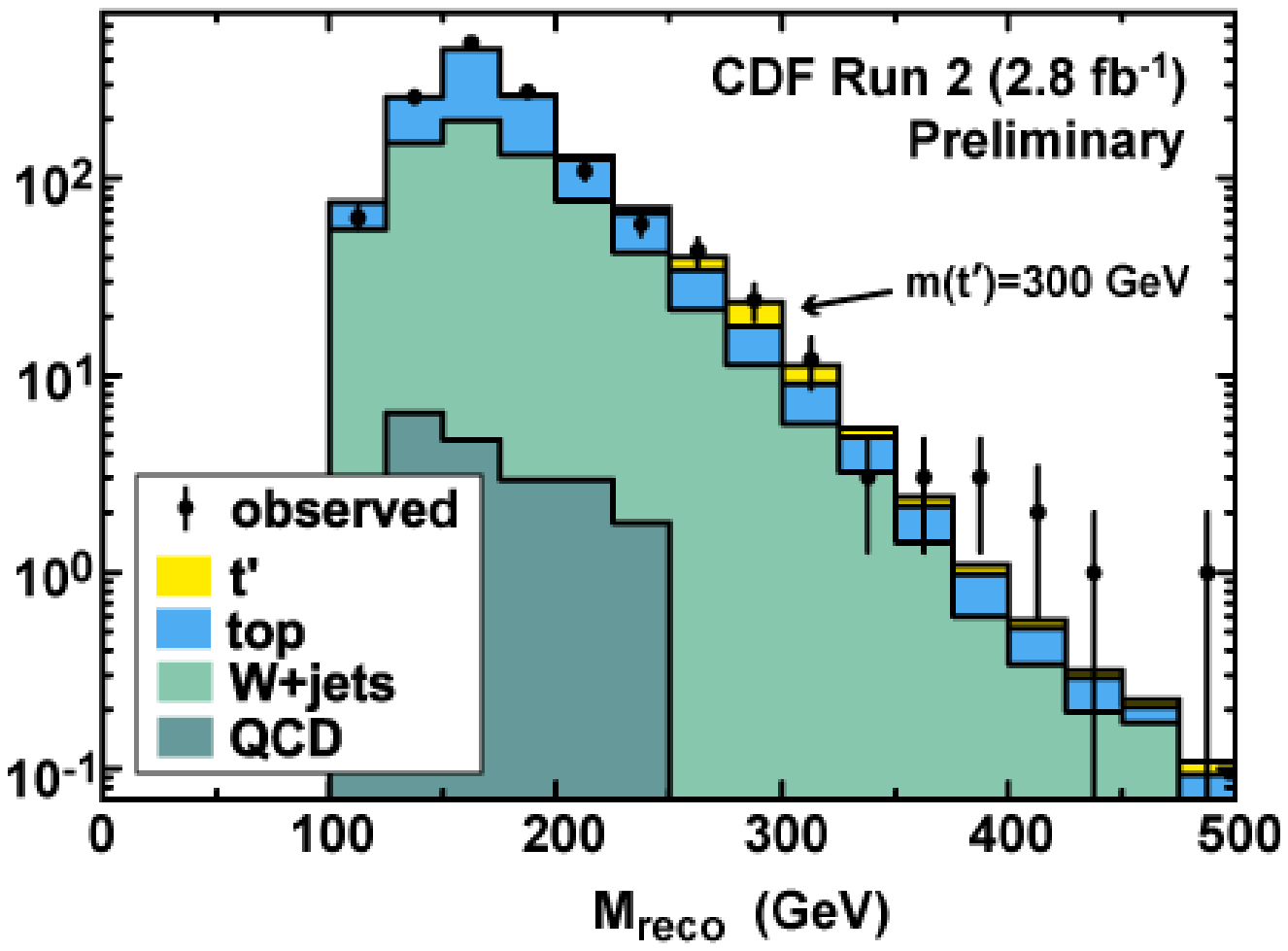}
    \caption{Fitted distribution of the reconstructed $t'$  mass in data and in the expected signal
      plus background. In this example a $t'$ mass of 300~\GeV is assumed.}
    \label{fig:CDF_tprime1}
  \end{minipage}
  \hfill
  \begin{minipage}[t]{0.48\linewidth}
    \centering
    \includegraphics[width=0.8\linewidth]{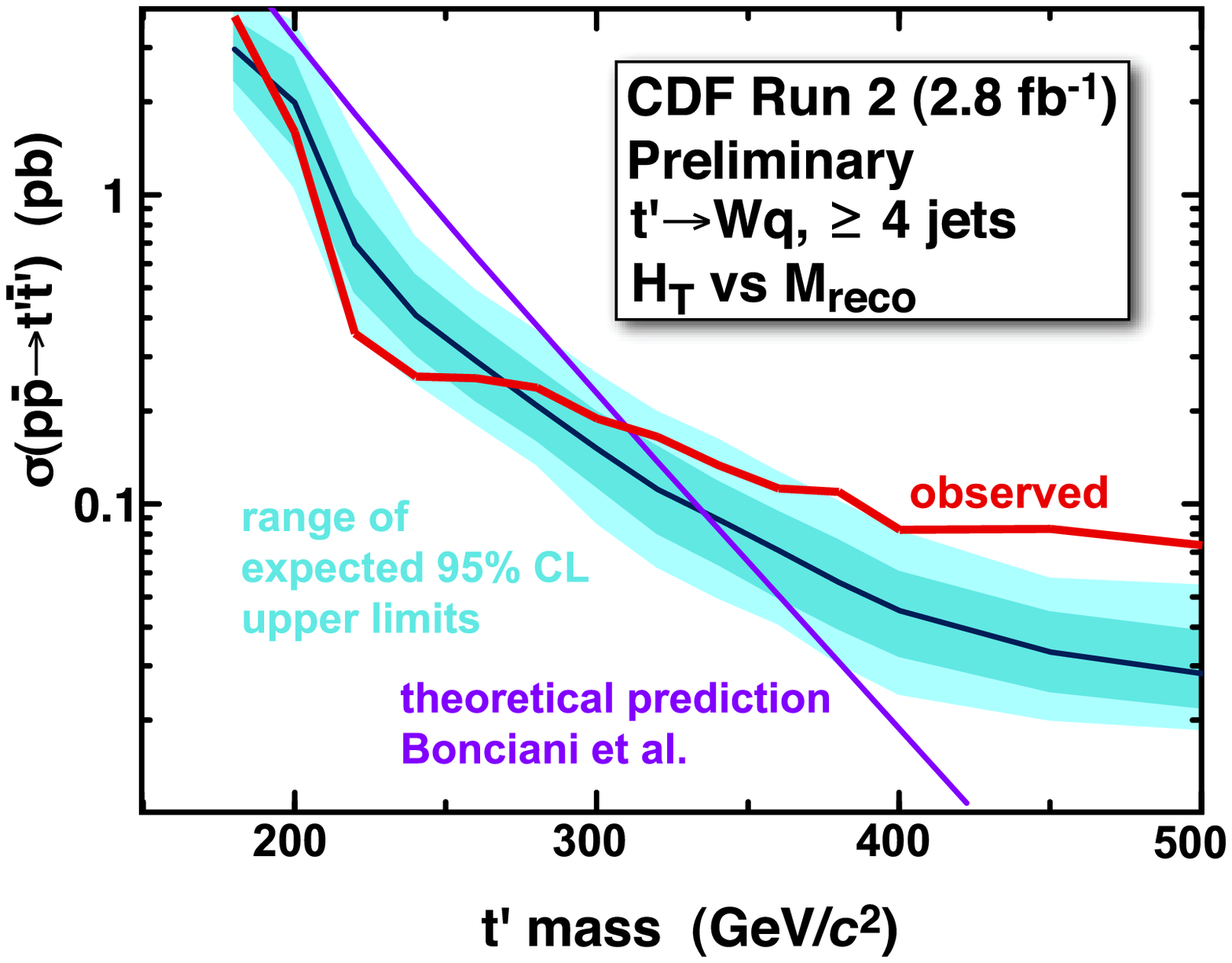}
    \caption{Expected and observed 95\% CL cross section limits on $t'$
    production as a function of the $t'$ mass.}
    \label{fig:CDF_tprime2}
  \end{minipage}
\end{figure}

\section{TOP-QUARK PAIRS AND THE HIGGS BOSON}
The production of a SM Higgs boson in association with a top quark
pair will be the only direct way to measure the top Yukawa coupling at
the LHC. In the mass region $m_H \lesssim 135~\GeV$ the dominant Higgs
boson decay mode is $H \to b\bar{b}$ which in the $\ell + \text{jets}$
final state for the top quark pair leads to the challenging signature
of one high $p_T$ lepton, large missing transverse energy and up to 6
jets of which four are $b$ jets. The \ttH cross section at the
Tevatron is very small (about 7~fb for $m_H = 115~\GeV$) and
observation of a Higgs boson in this channel alone is not possible.
Nevertheless, the \ttH channel can contribute to the total discovery
or exclusion potential of the Tevatron. \DO has searched for $\ttH \to
\ell\nu bjjbbb$ in 2.1~\ifb of data~\cite{c:ttH}. To increase the
sensitivity, the analysis has been divided into events with four and
at least five jets and with one, two, or three $b$-tagged jets. The
distribution of the scalar sum of the transverse momentum of the up to
five leading jets, $H_T$, shows good discrimination between the \ttH
signal and the backgrounds and is used in the limit setting procedure.
The largest signal contribution is expected from events with five jets
and at least three $b$-tagged jets. The combination of all channels
leads to a limit on the cross section relative to the SM expectation
as shown in Fig.~\ref{fig:D0_ttH2}. Additional tests show that the
observed and expected cross section limits agree within one standard
deviation. For $m_H = 115~\GeV$ a cross section greater than 64 times
the SM value can be excluded.

\begin{figure}
  \begin{minipage}[t]{0.48\linewidth}
    \centering
    \includegraphics[width=0.8\linewidth]{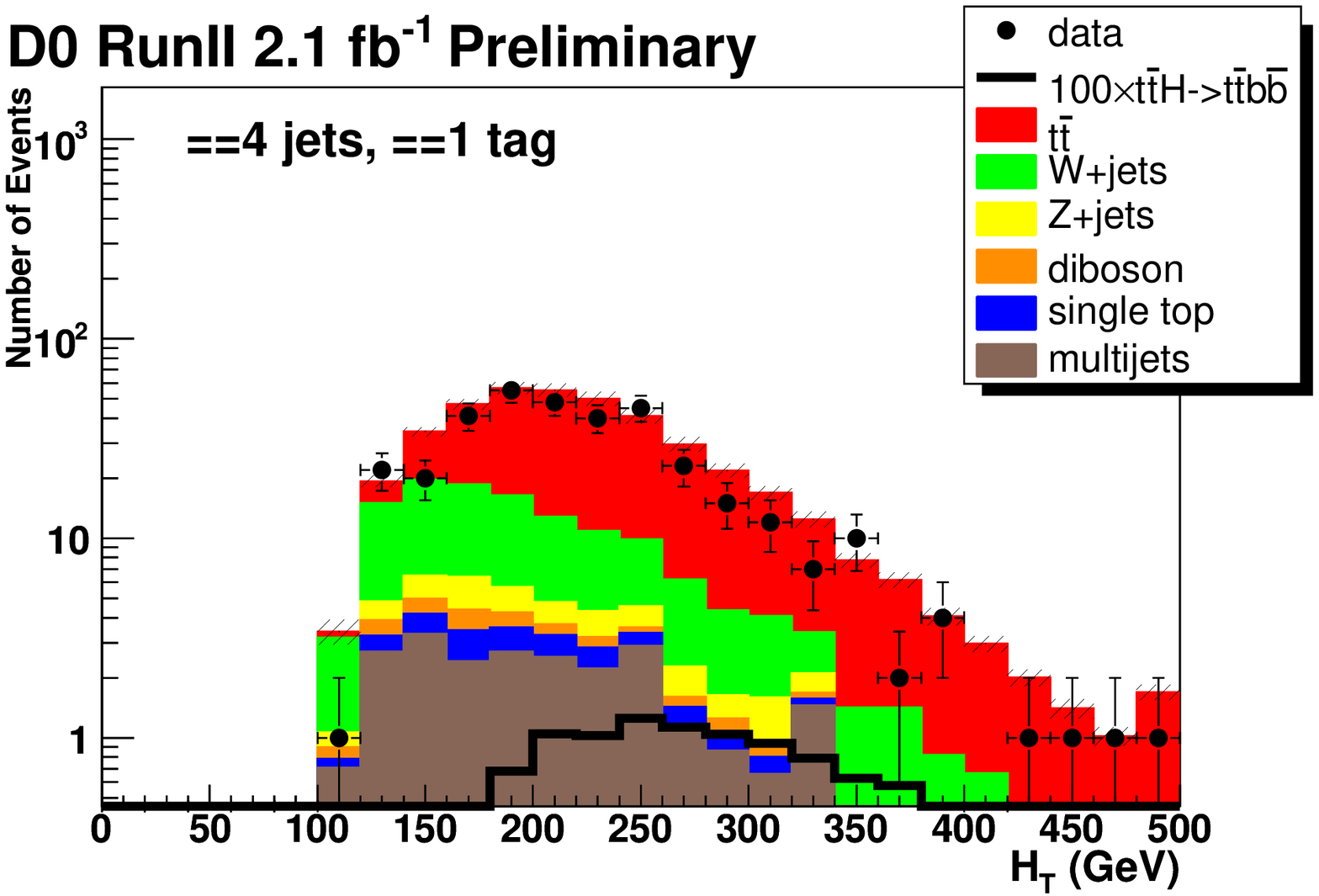}
    \caption{The $H_T$ distribution in data, expected backgrounds and
    the \ttH signal (multiplied by a factor of 100) for events with
    four jets and one $b$-tagged jet.)}
    \label{fig:D0_ttH1}
  \end{minipage}
  \hfill
  \begin{minipage}[t]{0.48\linewidth}
    \centering
    \includegraphics[width=0.6\linewidth]{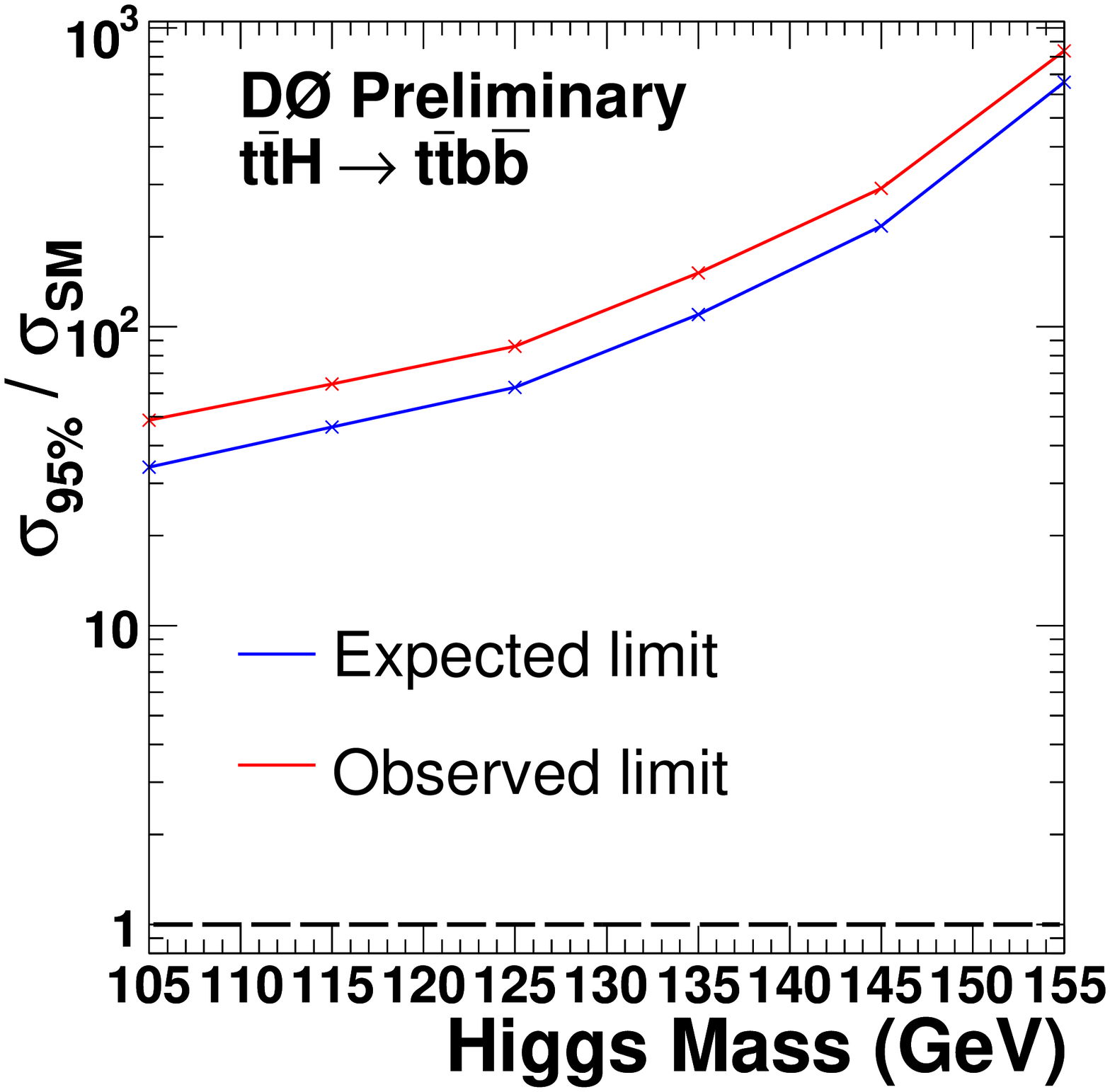}
    \caption{Expected and observed 95\% CL cross section times
      branching ratio limits on the \ttH process relative to the
      SM expectation as a function of the Higgs boson
      mass.}
    \label{fig:D0_ttH2}
  \end{minipage}
\end{figure}




\begin{thebibliography}{9}   


\bibitem{c:CDF_Zprime1}
  T.~Aaltonen {\it et al.}  [CDF Collaboration],
  Phys.\ Rev.\  D {\bf 77}, 051102 (2008)
  [arXiv:0710.5335 [hep-ex]].

\bibitem{c:D0_Zprime} 
  D0 Note 5600-CONF, \url{http://www-d0.fnal.gov/Run2Physics/WWW/results/prelim/TOP/T65/}.

\bibitem{c:CDF_Zprime2} 
  CDF Conference Note 9164, \url{http://www-cdf.fnal.gov/physics/new/top/confNotes/cdf9164_MG_DLM.pdf}.

\bibitem{c:DLM} CDF DLM
  A.~Abulencia {\it et al.}  [CDF Collaboration],
  Phys.\ Rev.\  D {\bf 73}, 092002 (2006)
  [arXiv:hep-ex/0512009].

\bibitem{c:CDF_stop} CDF Conference Note 9439,
  \url{http://www-cdf.fnal.gov/physics/new/top/2008/tprop/Stop/images2_7InvFb/stopDilCdfNote_2700InvPb.pdf}.

\bibitem{c:CDF_NW} 
  A.~Abulencia {\it et al.}  [CDF Collaboration],
  Phys.\ Rev.\ D {\bf 73}, 112006 (2006) [arXiv:hep-ex/0602008].

\bibitem{c:D0_stop} 
  D0 Note 5438-CONF, \url{http://www-d0.fnal.gov/Run2Physics/WWW/results/prelim/TOP/T52/}.

\bibitem{c:tprime} 
  CDF Conference Note 9446, \url{http://www-cdf.fnal.gov/physics/new/top/confNotes/cdf9446_tprime_public_2.8.pdf}.

\bibitem{c:ttH} 
  D0 Note 5739-CONF, \url{http://www-d0.fnal.gov/Run2Physics/WWW/results/prelim/HIGGS/H58/}.

\end{thebibliography}
\end{document}